\documentclass{elsarticle}

\usepackage{graphicx}
\usepackage{epsfig}

\begin{document}

\begin{frontmatter}

\title{Understanding how both the partitions of a bipartite network affect its one-mode projection}

\author[isi]{Animesh Mukherjee}

\author[msr]{Monojit Choudhury}

\author[kgp]{Niloy Ganguly}

\address[isi]{Institute for Scientific Interchange (ISI), Viale Settimio Severo 65, 10133 Torino, Italy}
\address[msr]{Microsoft Research India, Bangalore -- 560080}
\address[kgp]{Department of Computer Science and Engineering, Indian Institute of Technology, Kharagpur, India -- 
721302}

\date{\today}

\begin{abstract}
It is a well-known fact that the degree distribution (DD) of the nodes in a partition of a bipartite network influences the DD of its one-mode projection on that partition. However, there are no studies exploring the effect of the DD of the other partition on the one-mode projection. In this article, we show that the DD of the other partition, in fact, has a very strong influence on the DD of the one-mode projection. We establish this fact by deriving the exact or approximate closed-forms of the DD of the one-mode projection through the application of generating function formalism followed by the method of iterative convolution. The results are cross-validated through appropriate simulations.
\end{abstract}

\end{frontmatter}


\section{Introduction}\label{intro}
A bipartite network consists of two partitions of nodes, say $U$ and $V$, such that edges connect nodes from different partitions, but never those in the same partition. A {\em one-mode projection} of such a bipartite network onto $U$ is a network consisting of the nodes in $U$; two nodes $u$ and $u'$ are connected in the one-mode projection, if and only if there exist a node $v \in V$ such that $(u,v)$ and $(u',v)$ are edges in the corresponding bipartite network. Many real-life networks are, in fact, one-mode projections of a more fundamental bipartite structure~\cite{Guillaume:04,Guillaume:06}. As an example, consider the friendship and word co-occurrence networks. The former arises from the underlying bipartite relationship of the individual to different places (pubs, family, workplace etc.) because friendship groups evolve around certain social contexts (e.g., people regularly meeting in a pub, or colleagues at a workplace). The latter arise from an underlying word-sentence bipartite network. Therefore, for several real-world complex systems as described in~\cite{Holme:03,Lambiotte:05b,Lind:05,Estrada:05}, understanding the underlying bipartite process turns out to be extremely important. 

In this bipartite process, there are precisely two components -- (a) the attachment process i.e., how ties get formed between individuals or words (we shall refer to this partition as $U$) and different entities like pubs, workplaces or sentences (partition $V$), and (b) the size distribution of the entities in $V$, for instance, the number of words in a sentence or number of individuals at a workplace. The effect of the former is heavily studied in the literature and it is well-known that the attachment in real world bipartite networks is largely preferential in nature~\cite{Newman:01,Ramasco:04,Peruani:07,Choudhury:08}. Nevertheless, the latter has not received much attention in the network community, even though the basic framework for computing the DD of the one-mode projection has been formulated long back in~\cite{NewmanWatts:01}. The popular but unrealistic assumption that the degree of the nodes in partition $V$ is a constant results in networks whose one-mode projection onto $U$ has a DD qualitatively identical to that of $U$ in the bipartite network. Here we show that under a more realistic assumption where the degrees of the nodes in $V$ are sampled from a distribution (which is not a constant), the DD of this one-mode projection is remarkably different from that of the DD of $U$. Our analysis reveals that the dependence of the DD of the one-mode projection on the DD of $V$ is so strong that even slight relaxation of the ``constant degree'' assumption, for instance if the DD of $V$ is peaked (normal and exponential distributions), leads to significantly different results.   

The {\em generating function} (GF) formalism introduced in~\cite{NewmanWatts:01} presents open equations of the one-mode DD and therefore it is difficult to derive a meaningful insight from these equations. The main contribution of this work lies in the derivation of the closed-forms for the DD of the one-mode projection under some realistic assumptions. We used the process of {\em iterative convolution} to arrive at our results, which enabled us to analytically study the influence of the DD of the partition $V$, so long overlooked in the literature. The results have been cross-validated through appropriate simulations. 

\section{Analysis of the degree distribution of the one-mode projection}\label{sec2}

Formally, the one-mode projection considered here is a graph where $u_i, u_j \in U$ are connected by an edge if there exists a node $v \in V$ such that there is an edge between (a) $u_i$ and $v$ and (b) $u_j$ and $v$ in the bipartite network. If there are $w$ such nodes in $V$ which are connected to both $u_i$ and $u_j$ in the bipartite network, then there are $w$ edges linking $u_i$ and $u_j$ in the one-mode projection. Alternatively, one can think of the one-mode projection as a weighted graph, where the weight of the edge $(u_i,u_j)$ is $w$. In the rest of the paper, we always consider the degree distribution of this weighted one-mode network.

Let us assume that the degree of the nodes in partition $V$ are sampled from a distribution $f_d$ with expected value $\mu$. Let us denote the degree and DD of a node $u \in U$ as $k$ and $p_k$ respectively in the bipartite network. Further, let $q$ denote the degree of the nodes in the one-mode projection on $U$. Let us call the probability that the node $u$ having degree $k$ in the bipartite network ends up as a node having degree $q$ in the one-mode projection $F_k$($q$). Also, let us denote the degree distribution of the nodes in $U$ in the one-mode projection by $p_u$($q$). If we assume that the degrees of 
the $k$ nodes in $V$ to which $u$ is connected to are $d_1$, $d_2$, $\dots$, $d_k$ then we can write
\begin{equation}\label{eq:q}
q = \sum_{i = 1 \dots k} (d_i-1)
\end{equation}
The probability that the node $u$ in the bipartite network is connected to a node in $V$ of degree $d_i$ is $d_if_{d_i}$, where $i = 1\dots k$. At this point, one might apply the GF formalism~\cite{NewmanWatts:01} to calculate the degree distribution of the nodes in the one-mode projection as follows. Let $f(x)=\sum_d f_d x^d$ denote the GF for the distribution of the node degrees in $V$, $p(x)=\sum_k p_k x^k$ denote the GF for the degree distribution of the nodes in $U$ and $g(x)$ denote the GF for $p_u$($q$) then it is straightforward to see from eq.~(70) of~\cite{NewmanWatts:01} that,
\begin{equation}\label{eq:newman}
g(x) = p(f'(x)/\mu)
\end{equation}

On suitable expansion of eq.~(\ref{eq:newman}) we obtain
\begin{equation}\label{eq:puq}
p_u(q) = \sum_{k} p_k F_k(q) 
\end{equation}
or,
\begin{equation}\label{eq:puqexp}
p_u(q) = \sum_{k} p_k \sum_{d_1 + d_2 + \dots + d_k - k = q} \frac{d_1 d_2 \dots d_k}{\mu^k} f_{d_1} f_{d_2} \dots f_{d_k}
\end{equation}

For peaked distributions we can make the assumption that there will be a finite probability only when $d \approx \mu$ and $\mu \gg 1$. Hence, $d_1+d_2$+$\dots+d_k \approx k\mu$ which implies that the arithmetic mean is roughly equal to the geometric mean. Therefore, we have $d_1 d_2$ $\dots d_k$ approximately equal to $\mu^k$. We shall shortly discuss in further details the bounds of this approximation (section~\ref{sec2partb}). However, prior to that, let us investigate, how this approximation helps in advancing our analysis. Under the 
assumption $\frac{d_1 d_2 \dots d_k}{\mu^k} = 1$, $F_k$($q$) can be thought of as the distribution of the sum of $k$ 
random variables each sampled from $f_d$. In other words, $F_k$($q$) tells us how the sum of the $k$ random variables 
is distributed if each of these individual random variables are drawn from the distribution $f_d$. This distribution 
of the sum can be obtained by the iterative convolution of $f_d$ for $k$ times\footnote{Apart from some special cases, $df_d$ is hard to convolve and so we work with the 
approximate $F_k$($q$) here.}. If the closed form expression for the 
convolution exists for a distribution, then we can obtain an analytical expression for $p_u(q)$. In the following, we 
shall attempt to find an expression for $p_u$($q$) assuming three different forms of the distribution $f_d$. As we 
shall see, $F_k$($q$) is different for each of these forms, thereby, making the degree distribution of the nodes in 
the one-mode sensitive to the choice of $f_d$. Since in the expression for $q$ (eq.~(\ref{eq:q})) we need to subtract one 
from each of the $d_i$ terms (i.e., each term is $(d_i-1)$ rather than $d_i$) therefore the mean of the distribution $F_k$($q$) has to be shifted accordingly.  

\subsection{Effect of the sampling distribution $f_d$}\label{sec2parta}

In this section, we shall analytically study the effect of the sampling distribution $f_d$ on the degree distribution 
of the one-mode projection of the bipartite network.   
 
\noindent {\em Delta function}: Let $f_d$ be a delta function of the form
\begin{equation}\label{eq:D1}
\delta{(d, \mu)} = \left\{ \begin{array}{ll}
1 & \textrm{if } d=\mu \\
0 & \textrm{otherwise} \\
\end{array}
\right.
\end{equation}
If this delta function is convolved $k$ times then the sum 
should be distributed as
\begin{equation}\label{eq:D2}
F_k(q) = \delta{(q, k\mu-k)} = \left\{ \begin{array}{ll}
1 & \textrm{if } q=k\mu-k\\
0 & \textrm{otherwise} \\
\end{array}
\right.
\end{equation}
Therefore, $p_u$($q$) exists only when $q=k(\mu-1)$ or $k = q/(\mu-1)$ and we have (also reported in~\cite{Choudhury:08})
\begin{equation}\label{eq:D3}
p_u(q) = \left\{ \begin{array}{ll}
p_k & \textrm{if } k=q/(\mu-1)\\
0 & \textrm{otherwise} \\
\end{array}
\right.
\end{equation} 
\noindent {\em Normal distribution}: If $f_d$ is a normal distribution of the form $N$($\mu$, $\sigma^2$) then the sum 
of $k$ random variables sampled from $f_d$ is again distributed as a normal distribution of the form $N$($k\mu$, 
$k\sigma^2$). Therefore, $\mathring{F}_k$($q$) is given by
\begin{equation}\label{eq:N1}
F_k(q) = N(k\mu-k, k\sigma^2)
\end{equation}
If we substitute the density function for $N$ we have
\begin{equation}\label{eq:N3}
p_u(q) = \frac{1}{\sigma\sqrt{2\pi}}\sum_{k} p_k k^{-0.5} \exp{\left(-\frac{(q-k(\mu-1))^2}{2k\sigma^2}\right)}
\end{equation} 
\noindent {\em Exponential distribution}: If $f_d$ is an exponential distribution of the form $E$($\lambda$) where 
$\lambda = 1/\mu$ then the sum of the $k$ random variables sampled from $f_d$ is known to take the form of a gamma 
distribution $\Gamma$($q$; $k$, $\mu$). Therefore, we have
\begin{equation}\label{eq:E1}
F_k(q) = \Gamma(q; k, \mu-1) 
\end{equation}
Thus, we have ($\lambda^{'} = 1/(\mu-1)$)
\begin{equation}\label{eq:E3}
p_u(q) = \lambda^{'}\sum_{k} p_k \frac{\exp{(-\lambda^{'}q)}(\lambda^{'}q)^{k-1}}{(k-1)!}
\end{equation}

\subsection{Choice of $p_k$ and illustration}\label{sect2partnull}

The framework presented above is applicable for any choice of $p_k$. Literature presents two broad categories of bipartite networks (a) where both partitions grow~\cite{Newman:01,Ramasco:04} and (b) where one partition is fixed~\cite{Peruani:07,Choudhury:08,Evans:07}. This second case is particularly interesting because it is appropriate to model discrete combinatorial systems (DCS)~\cite{pinker:94}. A DCS consists of a finite set of elementary units (e.g., codons and letters/phonemes, i.e., $U$) that serves as its basic building blocks and the system, in turn, is a collection of a potentially infinite number of discrete combinations of these units (e.g., genes and languages, i.e., $V$). In this case, briefly, the stochastic model used to construct the bipartite network is as follows: at each time step $t$, a new node is introduced in the set $V$ which preferentially connects itself to $\mu$ nodes in $U$. Let $v_t$ be the node added to $V$ during the $t^\textrm{th}$ time step. Let $\widetilde{A}(k_{i}^{t})$ denote the probability that a new node $v_t$ entering $V$ attaches itself to a node $u_i \in U$, where $k_{i}^{t}$ refers to the degree of the node $u_i$ at time step $t$. $\widetilde{A}(k_{i}^{t})$ defines the attachment kernel and takes the form 
\begin{equation}\label{eq:kernel_attachment}
\widetilde{A}(k_{i}^{t}) = \frac{\gamma k_{i}^{t} + 1}{\sum_{j=1}^N (\gamma k_{j}^{t} + 1)} \
\end{equation}
where the sum in the denominator runs over all the nodes in $U$, and $1/\gamma$ is the tunable model parameter which is usually referred to as the the {\em initial attractiveness}~\cite{Mendes:03}. Note that the higher the value of $1/\gamma$, the higher the randomness in the system.

\begin{figure}[!t]
\begin{center}
\includegraphics[width=4.5in,height=4.2in]{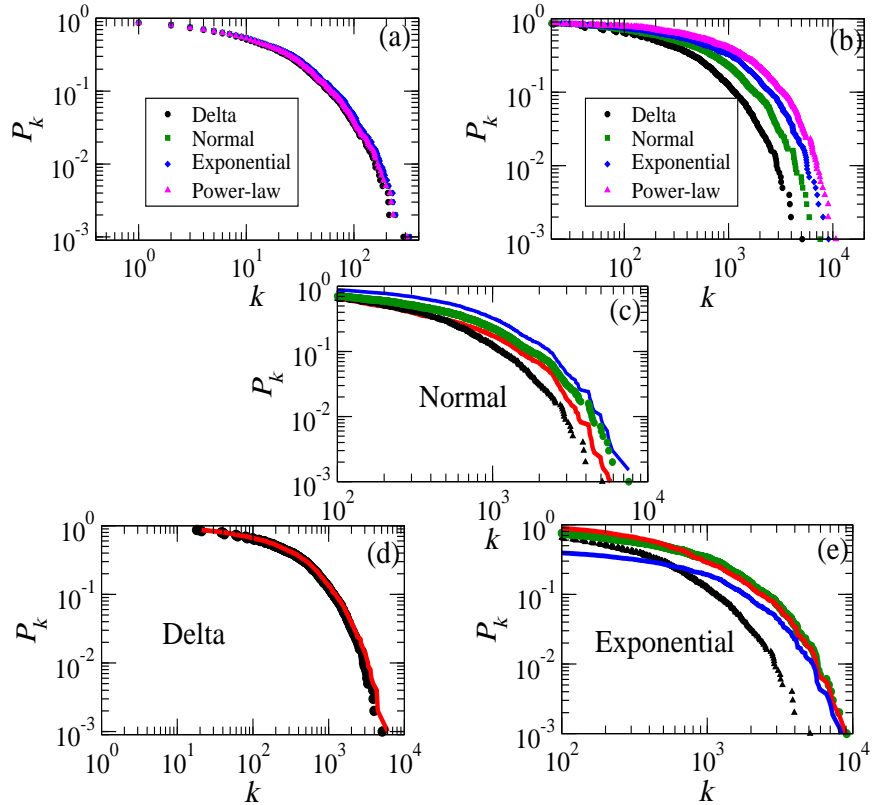}
\caption{Degree distribution of bipartite networks and the corresponding one-mode projections in doubly-logarithmic scale. $N=1000$, $t=1000$ and $\gamma=2$. For stochastic simulations, the results are averaged over 100 runs. All 
the results are appropriately normalized. (a) Degree distributions of the nodes in $U$ in the bipartite network generated through 
stochastic simulations when $f_d$ is a (i) normal ($\mu=22$, $\sigma=13$), (ii) delta ($\mu=22$), (iii) exponential ($\mu=\frac{1}{\lambda}=22$) 
and (iv) power-law (exponent $\lambda=1.16$, $\mu=22$, simulated within the interval $[k_{min}=1, k_{max}=311]$) distribution; (b) the degree distributions of the one-mode 
projections of the bipartite networks in (a); (c) match between stochastic simulations (green circles) and eq.~(\ref{eq:N3}) (red line) with $\mu=22$, $\sigma=13$; black triangles indicate the case where $f_d$ is a constant with $\mu=22$; blue line shows how the result deteriorates when $\sigma$ is 100 times larger; 
(d) match between stochastic simulations (black circles) and eq.~(\ref{eq:D3}) (red line) where $\mu=22$; (e) match 
between stochastic simulations (green circles) and eq.~(\ref{eq:E3}) (red line) where $\mu=22$; 
blue lines show the plot for eq.~(\ref{eq:expclose}); black triangles indicate the case where $f_d$ is a constant with $\mu=22$ (given as a reference to show that even the approximate eq.~(\ref{eq:expclose}) produces better results).}  
\label{eplfig1}
\end{center}
\end{figure}

~\cite{Peruani:07,Choudhury:08} shows that the emergent  $p_k$ for the above model asymptotically approaches a $\beta$-distribution such that $p_k =  M(k/t)^{\gamma^{-1}-1}(1 - k/t)^{\eta - \gamma^{-1}-1}$ where $\eta = N/\mu\gamma$ and $M$ is a normalization constant. Note that $\beta$-distributions are more general than power-law distributions (noticed in expanding bipartite networks~\cite{Newman:01,Ramasco:04}) since they can take different forms ranging from a normal distribution to a heavy-tailed distribution depending on the two parameters of the distribution. Therefore, we would illustrate the results of the equations with this type of a $\beta$-distribution presented in~\cite{Peruani:07,Choudhury:08}. Figure~\ref{eplfig1}(a) shows the cumulative degree distribution of the nodes in $U$ in the bipartite network assuming that nodes in $V$ arrive with degrees sampled from $f_d$ which can take the form of a (i) normal, (ii) delta, (iii) exponential and (iv) power-law distribution each with mean ($\mu=22$). Note that we use the probability mass functions rather than the probability density functions (as in the theoretical analysis) for the simulation results reported in this figure. Further, note that the 
standard deviation ($\sigma$) of the normal distribution is controlled in such a way that the value of the random 
variable $d$ is never negative. Figure~\ref{eplfig1}(b) shows the degree distributions of the one-mode projections 
corresponding to the bipartite networks generated for Figure~\ref{eplfig1}(a). The result clearly implies that the degree 
distribution of the one-mode projection varies depending on how the degrees of the nodes in $V$ are distributed 
although the degree distribution remains unaffected for all the bipartite networks generated. Figure~\ref{eplfig1}(c)--(e) shows 
the match of the analytical expressions (with appropriate normalization) derived with the respective stochastic 
simulations. Note that if $f_d$ is power-law distributed, the standard deviation $\sigma$ diverges and therefore an analytical study of this case is beyond the scope of the paper. In addition, no clear closed form solution for the convolution exists for this case. However, the stochastic simulation (Figure~\ref{eplfig1}(b)) indicates that this choice results in an one-mode degree distribution that is quite different from the case where $f_d$ is constant. 

\subsection{Approximation bounds}\label{sec2partb}

Here we discuss the limitations of the approximation that we made in eq.~(\ref{eq:puqexp}) by assuming that $\frac{d_1 d_2 \dots d_k}{\mu^k} = 1$. We shall employ the GF formalism to find the necessary condition (in the asymptotic limits) for our approximation to 
hold. More precisely, we shall attempt to estimate the difference in the means (or the first moments) of the exact and 
the approximate expressions for $p_u(q)$ and discuss when this difference is negligible which in turn serves as a 
necessary condition for the approximation to be valid. We shall denote the generating function for the approximate 
expression of $p_u(q)$ as $g_{app}(x)$. In this case, the GF encoding the probability that the node $u$ is 
connected to a node in $V$ of degree $d$ is simply $\sum_{d-1} f_{d-1}x^{d-1}$ which is $f(x)/x$ and consequently, $F_k(q)$ is given by 
$(f(x)/x)^k$. Therefore,
\begin{equation}\label{eq:genappr}
g_{app}(x) = \sum_k p_k \left[\frac{f(x)}{x}\right]^k = p(f(x)/x)
\end{equation}
Now we can calculate the first moments for the approximate and the exact $p_u(q)$ by evaluating the derivatives of  
$g_{app}(x)$ and $g(x)$ respectively at $x=1$. We have
\begin{equation}\label{eq:gapp}
g_{app}'(1) = \frac{d}{dx}p(f(x)/x)|_{x=1} = (t/N)\mu(\mu-1)
\end{equation}
Similarly,
\begin{equation}\label{eq:gexc}
g'(1) = \frac{d}{dx}p(f'(x)/\mu)|_{x=1} = (t/N)\mu(\mu-1) + (t/N)\sigma^2
\end{equation}

Thus, the mean of the approximate $p_u(q)$ is smaller than the actual mean by $(t/N)\sigma^2$. Clearly, for 
$\sigma=0$, the approximation gives us the exact solution, which is indeed the case for delta functions. Also, in the 
asymptotic limits, if $\sigma^2 \ll N$ (with a scaling of $1/t$), the approximation holds good. However, as the value 
of $\sigma$ increases the results start deteriorating (blue line in Figure~\ref{eplfig1}(c)).

\subsection{Closed-form expression}\label{sec2partc}

Finally, it remains to be mentioned that in some special cases it is possible to derive a closed form expression for $p_u(q)$. If $p_k F_k(q)$ takes up a very simple form then a closed form expression for $p_u(q)$ can be derived straight away. For instance, if in eq.~(\ref{eq:E3}), $p_k \propto (k-1)!$, then one can easily show by changing the discrete sum to a continuous integral that
\begin{equation}\label{eq:simpclose}
p_u(q) = \frac{\lambda^{'}\exp{(-\lambda^{'}q)}}{\ln(\lambda^{'}q)} [(\lambda^{'}q)^{k-1}-(\lambda^{'}q)^{-1}]
\end{equation}
There can be a second situation too. One can think of $p_k F_k(q)$ as a function $F$ in $q$ and $k$, i.e., $p_k F_k(q) = F(q,k)$. If
$F(q,k)$ can be exactly (or approximately) factored into a form like $\widehat F(q) \widetilde F(k)$ then $p_u(q)$ 
becomes
\begin{equation}\label{eq:closedform}
p_u(q) = \widehat F(q) \sum_k \widetilde F(k) 
\end{equation}   
Changing the sum in eq.~(\ref{eq:closedform}) to its continuous form we have
\begin{equation}\label{eq:closedform1}
p_u(q) = \widehat F(q) \int^{\infty}_{0} \widetilde F(k)dk = A\widehat F(q) 
\end{equation}   
where $A$ is a constant. Thus, the nature of the resulting distribution is dominated by the function $\widehat F(q)$. For instance, in case of exponentially distributed $f_d$, with some algebraic manipulations and certain approximations\footnote{We use Stirling's approximation~\cite{Abramowitz:65} replacing $(k-1)!$ in the denominator of eq.~(\ref{eq:E3}) by $\sqrt{2\pi(k-1)}\left(\frac{k-1}{e}\right)^{k-1}$. Further, we assume $\left(\frac{q}{(\mu-1)(k-1)}\right)^{k-1} \rightarrow 1$ in the asymptotic limits. This makes the rest of the derivation possible.} one can show that (blue line in Figure~\ref{eplfig1}(e))
\begin{equation}\label{eq:expclose} 
p_u(q) \approx A{\bf EXP}\left(q;\frac{1}{\mu-1}\right)
\end{equation}
where {\bf EXP}() is the exponential distribution function.

\section{Discussion}\label{conc}

In this paper, we identified that the degree distribution of the one-mode projection of a bipartite network onto the partition $U$ is sensitive to the degree distribution of the other partition $V$. Further, we showed that if partition $V$ corresponds to a peaked distribution then it is possible to derive closed form expression for the one-mode degree distribution. The derivation of the closed form solution for the one-mode degree distribution points to the fact that this distribution is {\em not always reminiscent} of $p_k$ (i.e., the degree distribution of the nodes in $U$ in the bipartite network) as has been demonstrated in the literature. While eq.~(\ref{eq:simpclose}) shows that this distribution could be a complex coupling of the terms $k$ and $q$, eq.~(\ref{eq:expclose}) shows that it might be completely dominated by $f_d$ (i.e., the distribution of the node degrees in $V$ in the bipartite network). We believe that this observation is an important departure from what have been reported so long in the literature. In addition, from our simulation results it is clear that the one-mode degree distribution is affected when the partition $V$ is not peaked (see the power-law case in Figure~\ref{eplfig1}(b) and the normal distribution case with high $\sigma$ in Figure~\ref{eplfig1}(c)). These results indicate that as the standard deviation $\sigma$ becomes more and more arbitrary the effect on the one-mode degree distribution is more and more pronounced. Hence, an important future attempt would be to analytically solve for cases where $f_d$ is not peaked, i.e., has arbitrary $\mu$, $\sigma$. A final interesting and non-trivial direction could be to perform a similar analysis as done here but limited to the unweighted versions of the one-mode networks.

\bibliographystyle{unsrt}
\bibliography{references}

\end{document}